\documentclass[11pt,a4paper]{emulateapj}
\usepackage{epsfig}
\usepackage{amsmath}
\usepackage{natbib}
\usepackage{color}
\usepackage{wasysym}

\begin{document}
\title{Effective induction heating around strongly magnetized stars}
\author{K.G.~Kislyakova$^{1,2}$, L.~Fossati$^{2}$, C.P.~Johnstone$^{1}$, L.~Noack$^{3}$, T.~L\"{u}ftinger$^{1}$, V.V.~Zaitsev$^{4}$, \& H.~Lammer$^{2}$}
\affil{$^{1}$University of Vienna, Department of Astrophysics, Vienna, Austria; kristina.kislyakova@univie.ac.at\\
$^{2}$Space Research Institute, Austrian Academy of Sciences, Graz, Austria\\
$^{3}$Freie Universit\"{a}t Berlin, Berlin, Germany\\
$^{4}$Institute of Applied Physics, Russian Academy of Sciences, Nizhny Novgorod, Russia
}
\date{\today}
\begin{abstract}

Planets that are embedded in the changing magnetic fields of their host stars can experience significant induction heating in their interiors caused by the planet's orbital motion. For induction heating to be substantial, the planetary orbit has to be inclined with respect to the stellar rotation and dipole axes. Using WX~UMa, for which the rotation and magnetic axes are aligned, as an example, we show that for close-in planets on inclined orbits, induction heating can be stronger than the tidal heating occurring inside Jupiter's satellite Io; namely, it can generate a surface heat flux exceeding 2\,W\,m$^{-2}$. An internal heating source of such magnitude can lead to extreme volcanic activity on the planet's surface, possibly also to internal local magma oceans, and to the formation of a plasma torus around the star aligned with the planetary orbit. A strongly volcanically active planet would eject into space mostly SO$_2$, which would then dissociate into oxygen and sulphur atoms. Young planets would also eject CO$_2$. Oxygen would therefore be the major component of the torus. If the O{\sc i} column density of the torus exceeds $\approx$10$^{12}$\,cm$^{-2}$, the torus could be revealed by detecting absorption signatures at the position of the strong far-ultraviolet O{\sc i} triplet at about 1304\,\AA. We estimate that this condition is satisfied if the O{\sc i} atoms in the torus escape the system at a velocity smaller than 1--10\,km\,s$^{-1}$. These estimates are valid also for a tidally heated planet.

\end{abstract}

\section{Introduction}

Low-mass M~dwarfs are the most plentiful stars in the Universe and often host small, rocky planets \citep{Dressing13}. Several fully convective M~dwarfs have been observed to have strong magnetic fields (kG strength), often with simple geometries (e.g., \citealp{Morin11,Lang12,Shulyak17}). Recently, \citet{K17} (hereafter K17) estimated the amount of induction heating caused by the star's rotation and the planet's orbital motion for the planets in the TRAPPIST-1 system \citep{Gillon17}. K17 showed that under some conditions, induction heating can significantly increase outgassing from planetary interiors and even lead to subsurface magma oceans for some mantle parameters. Here, we further develop the model of K17 and apply it to a rocky planet orbiting WX~UMa, a late M~dwarf with a particularly strong dipole-dominated magnetic field with a strength of 7.3\,kG \citep{Shulyak17}. Unlike the model applied by K17, we assume that the stellar rotation and the magnetic dipole axes are aligned, as was found for WX~UMa. In this case, induction heating arises only if the planetary orbital plane is inclined with respect to the stellar magnetic axis. Under these conditions, the planet experiences constant changes of the magnetic field strength along its orbital motion, i.e. the magnetic field is not varying due to stellar rotation (as considered in K17), but due to planet's orbit around the star. Since the planetary mantle is conductive, the varying magnetic field strength generates eddy currents in the mantle, which then dissipate as heat.

 
In addition to induction heating, energy release due to tidal friction can be a very powerful source inside exoplanets orbiting M~dwarfs (e.g., \citealp{Driscoll15}). For multi-planet systems, the eccentricities can be large for Gyrs, making tidal heating a likely dominant heating source in many multi-planet systems. For example, for the \mbox{TRAPPIST-1} system, tidal heating likely exceeds induction heating \citep{Luger17}. We show that for planets orbiting strongly magnetized M~dwarfs, induction heating can be an energy source more powerful than tidal heating, especially for single-planet systems in which the eccentricities are quickly damped, thus making tidal heating negligible.

A model of induction heating similar to K17 has previously been applied to hot Jupiters orbiting T~Tauri stars by \citet{Laine08}. They have shown that energy release due to induction heating powered by the stellar magnetic field and rotation can inflate the planet and lead to Roche lobe overflow of planetary material. Their work differs from ours in several ways: first of all, we consider induction effects in rocky planets and not in gaseous giants, second, we consider main-sequence stars, and third, we take into account the effects of the stellar wind on the stellar magnetic field.

This paper is organized as follows. Section~\ref{sec_model} describes the models and assumptions used in the manuscript. Section~\ref{sec_results} presents the results of our computations. Section~\ref{sec_obs} addresses the possible observability of signatures connected with the presence of induction heating. Finally, we gather our conclusions in Section~\ref{sec_consclusions}.

\pagebreak

\section{Model}
\label{sec_model}

\subsection{Magnetic fields of late M dwarfs. The case of WX~UMa.}

Late M~dwarfs have been observed to often host strong magnetic fields with simple dipolar geometries  \citep{Morin11,Lang12,Vidotto14}. Often the magnetic dipole axis is aligned with the rotation axis, although any angle between the two is possible \citep{Lang12,Morin10,See15,Fares13,Fares17}. Recently, \citet{Shulyak17} have shown that some late M~dwarfs host magnetic fields with a dipolar strength exceeding 5\,kG. An example is the M6.0 star WX~UMa, which has a mass of 0.1\,M$_\odot$, a radius of 0.12\,R$_\odot$, and hosts a magnetic field with a dipolar strength of 7.3\,kG. Zeeman Doppler Imaging (ZDI) shows that its magnetic field is strongly dipole-dominated and that the magnetic dipole axis nearly coincides with the stellar rotation axis \citep{Morin10}. Therefore, a planet orbiting WX~UMa on an inclined orbit would be embedded into a constantly varying magnetic field, with the period of the magnetic field variation equal to planet's orbital period.

\citet{Yadav15} developed a model for the dynamo of fully convective M~dwarfs capable of reproducing various commonly observed features, such as very strong dipole-dominated magnetic fields. They concluded that fast rotating stars develop stable and strong magnetic fields with the dipole axis aligned (within 20$^\circ$) with the stellar rotation axis (R.~Yadav, private communication). \citet{Yadav16} showed that as the rotation of late M~dwarfs decelerates with age, stars seem to switch to a different dynamo regime and develop magnetic cycles. According to their model, the switch happens at a rotational period of about 20\,days. WX~UMa has a rotational period of only 0.78\,days, which implies that it has a young age \citep{Vidotto14,Reiners12}, and indicates that its magnetic field strength should be very stable for at least a few Gyr. 

Since we assume induction heating is powered by the planetary orbital motion, the deceleration of the stellar rotation rate with time  does not play a major role. In general, lower mass stars such as M~dwarfs evolve and decelerate slower than higher mass stars (e.g., \citealp{West08,Johnstone15}).

To calculate the magnetic field at a given orbital distance, we employ a potential-field source-surface model (PFSS), which is commonly used to model stellar magnetic fields \citep{JohnstoneThesis}. The field is assumed to be potential (current-free) within the stellar corona, which extends from the stellar surface to the `source surface' at $R_{ss}$, and then radial further away from the star. This assumption approximates the effects of the stellar wind on the magnetic field structure, which are significant because, for a dipole configuration, when \mbox{$r < R_{ss}$}, the field strength decreases with $r^{-3}$, while the decrease goes as $r^{-2}$ at larger distances. The source surface radius is a free parameter that could lead to uncertainties in the magnetic field by a factor of a few. At the planetary orbit, the field strengths is proportional to $R_{ss}^{-3}$. Larger source surfaces lead to weaker magnetic fields, and vice versa. We assume $R_{ss}=2.5$ stellar radii in our calculations, which is based on observations and modeling of the solar magnetic field \citep{Arden14}, as well as for modeling of observed magnetic fields of M~dwarfs \citep{Lang12}.

\subsection{Architecture of exoplanetary systems and possible inclinations.}

According to observations, exoplanetary systems can have diverse architectures \citep{Winn15}. K17 considered the case of a co-planar system, where the orbital plane of the planets is perpendicular to the stellar rotation axis. In their model, magnetic field variations at the planet's location arise due to the inclination of the stellar dipole axis with respect to the stellar rotation axis. 
 
As discussed above, while magnetic dipoles that are inclined with respect to stellar rotation are possible, it is a rather rare case among late M~dwarfs. Induction heating can still play a role in such systems when the planet's orbit is inclined. Although highly inclined orbits are more often observed for planets orbiting G and F~dwarfs, they are possible also for planets orbiting M~dwarfs \citep{Winn15}. Recently, \citet{Bourrier18} showed that the eccentric orbit of GJ~436b orbiting an M2.5 star is nearly perpendicular to the star's equator, proving that high inclinations are possible also for planets orbiting M~dwarfs. 

We calculate induction heating for all orbital inclinations from 0$^\circ$ to 90$^\circ$. In our model, the varying magnetic field at the planet's orbit arises due to planet's orbital motion. The energy budget inside rocky planets depends on the formation phase (accretional and gravitational energies released during core formation), and heating sources including induction, tidal and radioactive heating. Here we concentrate only on induction heating efficiency. 

\subsection{Interior and induction heating model.}

\begin{figure*}
  \begin{center}
        \includegraphics[width=1.0\columnwidth]{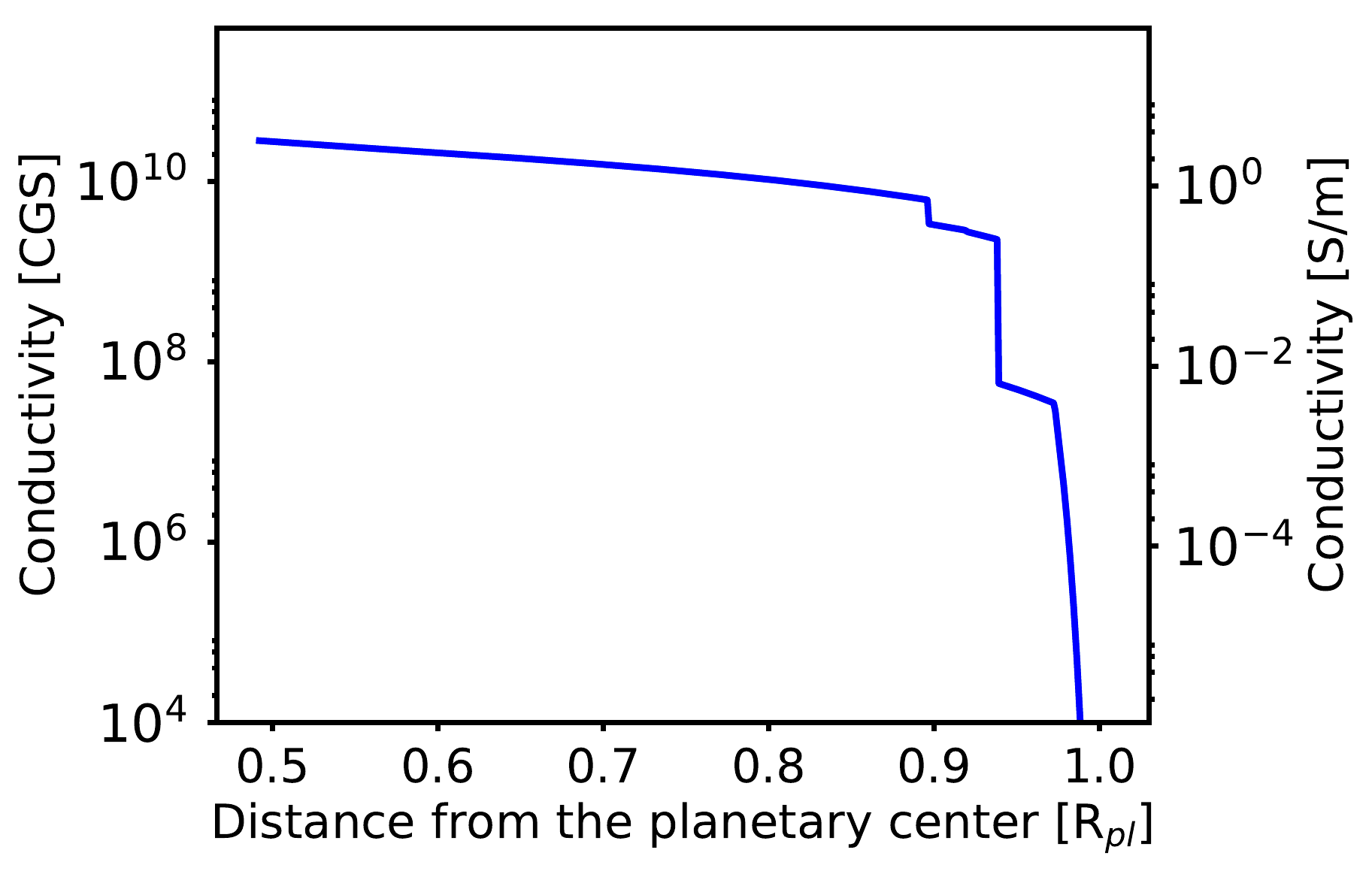}
        \includegraphics[width=1.0\columnwidth]{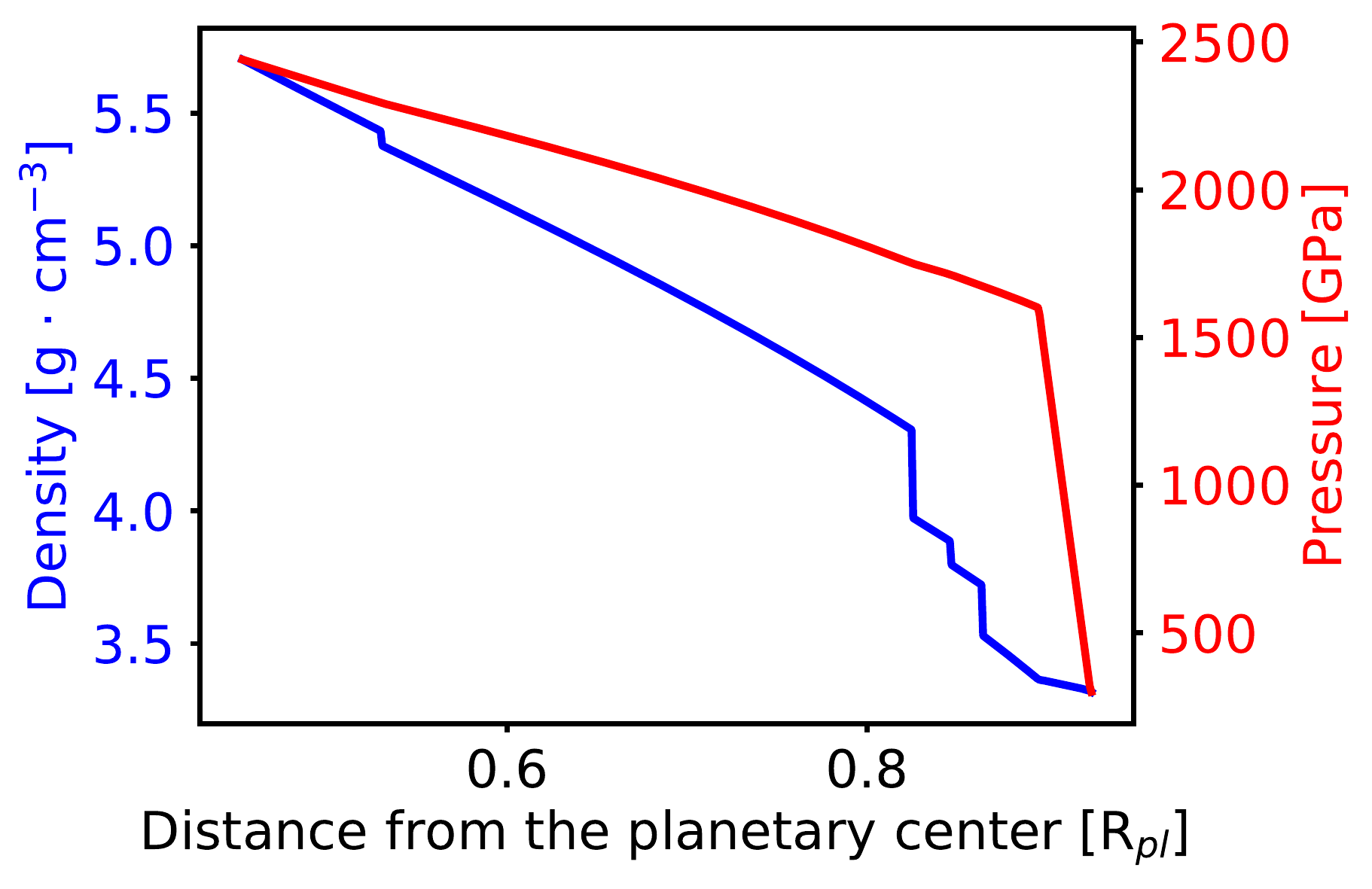}
        \caption{Left: Earth-like conductivity profile for the solid mantle calculated according to \citet{Yoshino08}, \citet{Yoshino13}, and \citet{Xu00}. The different phases in the mantle are marked by sharp changes in the conductivity profile. Right: Density and pressure profiles for an Earth-sized and Earth-massed planet \citep{Noack16CHIC}. }        
     \label{f_cond} 
  \end{center}
\end{figure*}

We calculate induction heating for an Earth-radius and Earth-mass stagnant lid planet orbiting WX~UMa. Since the properties a rocky exoplanet around WX~UMa would have are unknown, we assume an Earth-like composition and density and an Earth-like conductivity profile (Fig.~\ref{f_cond}). The density profile has been calculated with an interior structure model CHIC \citep{Noack16CHIC}. As K17, we assume a conductivity profile for a dry and iron-poor silicate mantle \citep{Xu00,Yoshino08,Yoshino13}. We also consider a terrestrial iron mass fraction of 35 wt-\%, which we assume accumulates entirely in the core. We additionally calculate induction heating for a fully molten mantle, for which we assume a fixed conductivity value throughout. Since molten rocks are known to have higher conductivities than the same minerals at temperatures below solidus, we assume a value of $5 \times 10^{10}$\,CGS ($\approx 5.6$\,Sm\,m$^{-1}$), which is higher than the conductivity of a non-molten mantle and is a reasonable approximation for molten rock (e.g., \citealp{Gaillard05}). 

In this article, we are primary interested in studying induction heating in close-in planets, which are likely airless bodies due to their proximity to the star and do not have a conducting layer of ionosphere surrounding them.
We calculate induction heating using the formalism developed by \citet{Parkinson83} and applied to exoplanets by K17. In this model, the planet is assumed to be a sphere made up of concentric layers; each layer has a uniform conductivity which is different for different layers. We solve the induction equation in every layer and calculate the magnetic field strength and current. Knowing the current and conductivity, we find the energy release within each layer (see \citet{Parkinson83} and K17, for details).

\section{Results}
\label{sec_results}

\begin{figure*}[t]
  \begin{center}
        \includegraphics[width=1.0\columnwidth]{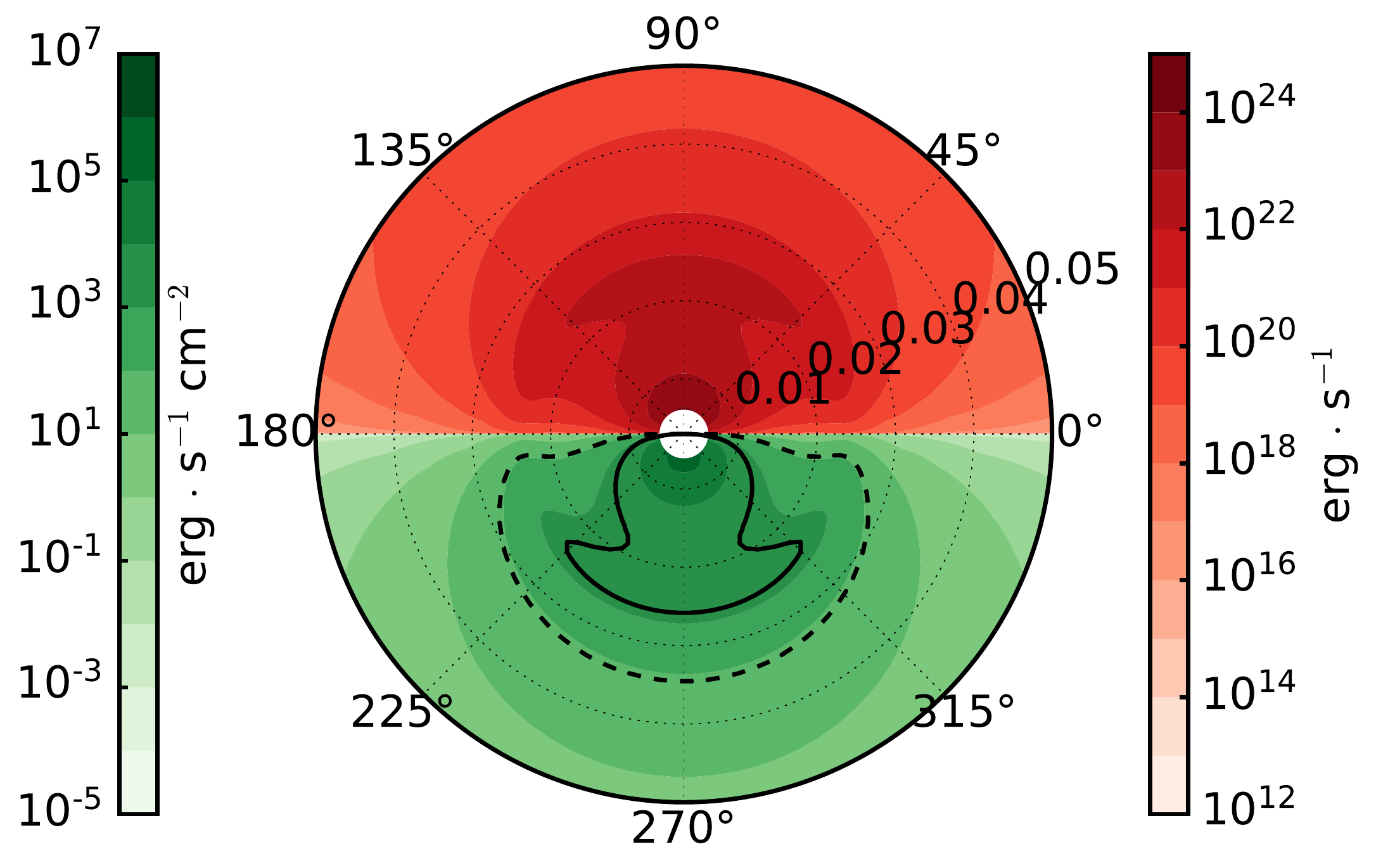}
        \includegraphics[width=1.0\columnwidth]{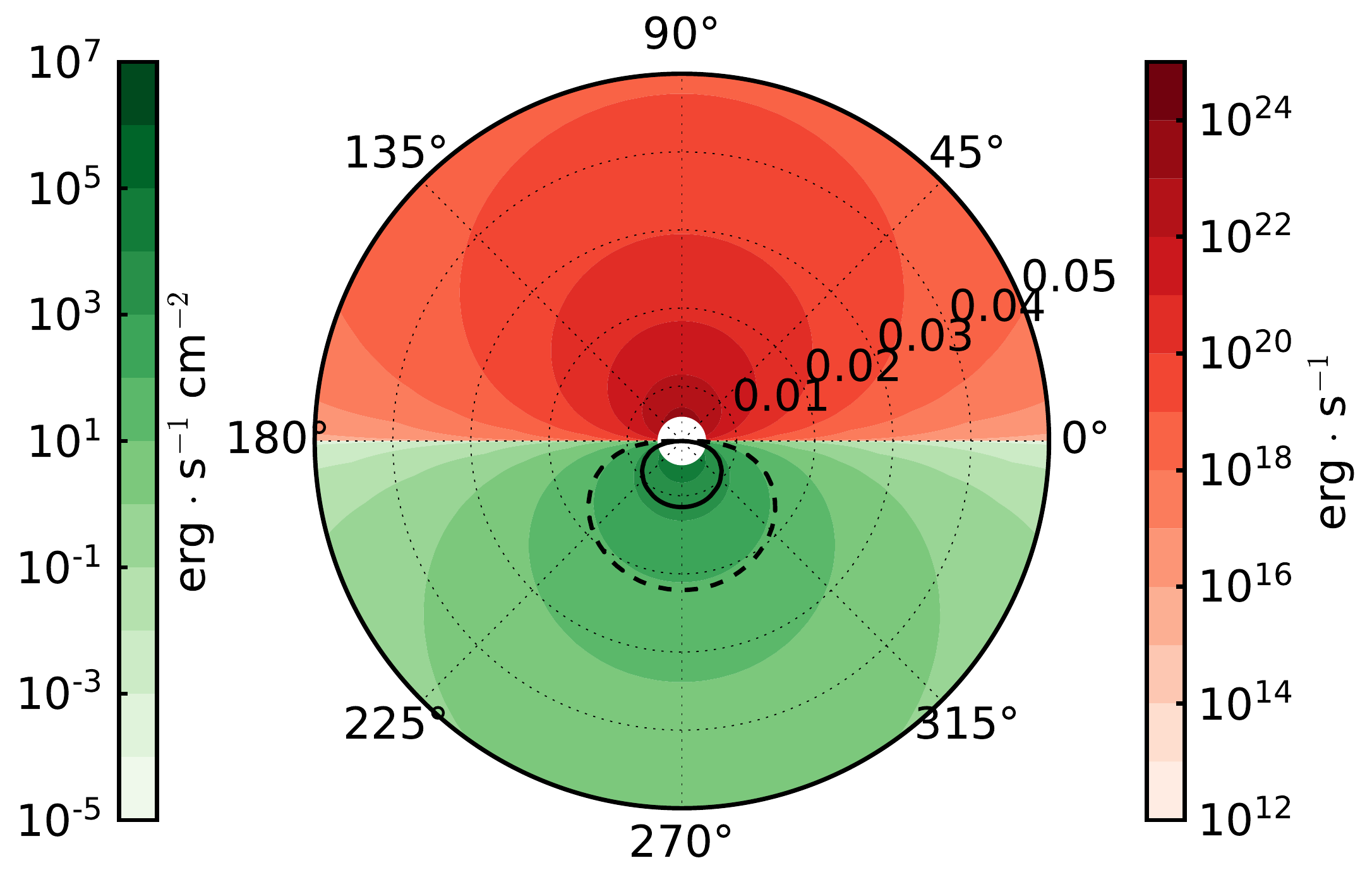}
        \includegraphics[width=1.0\columnwidth]{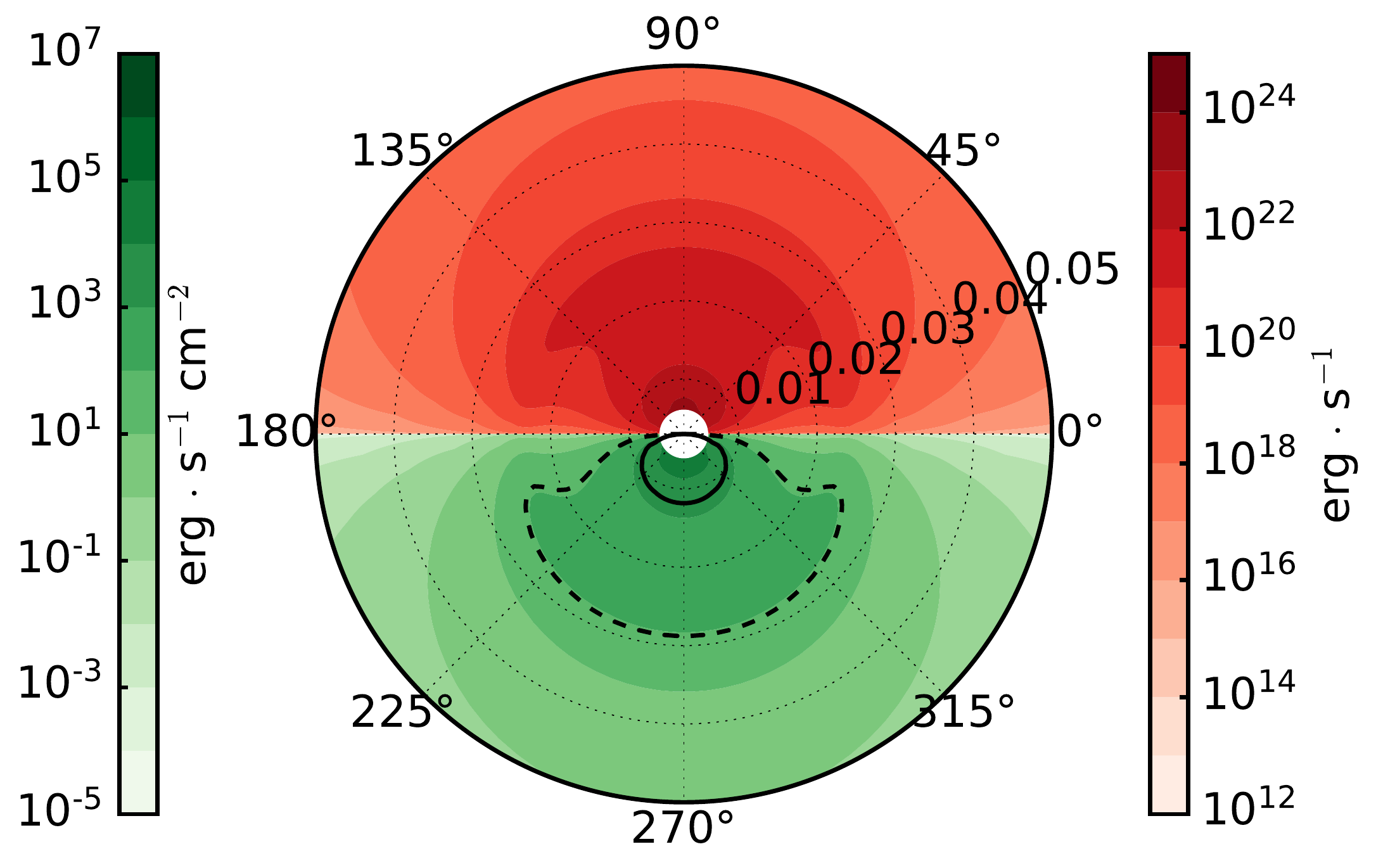}
        \includegraphics[width=1.0\columnwidth]{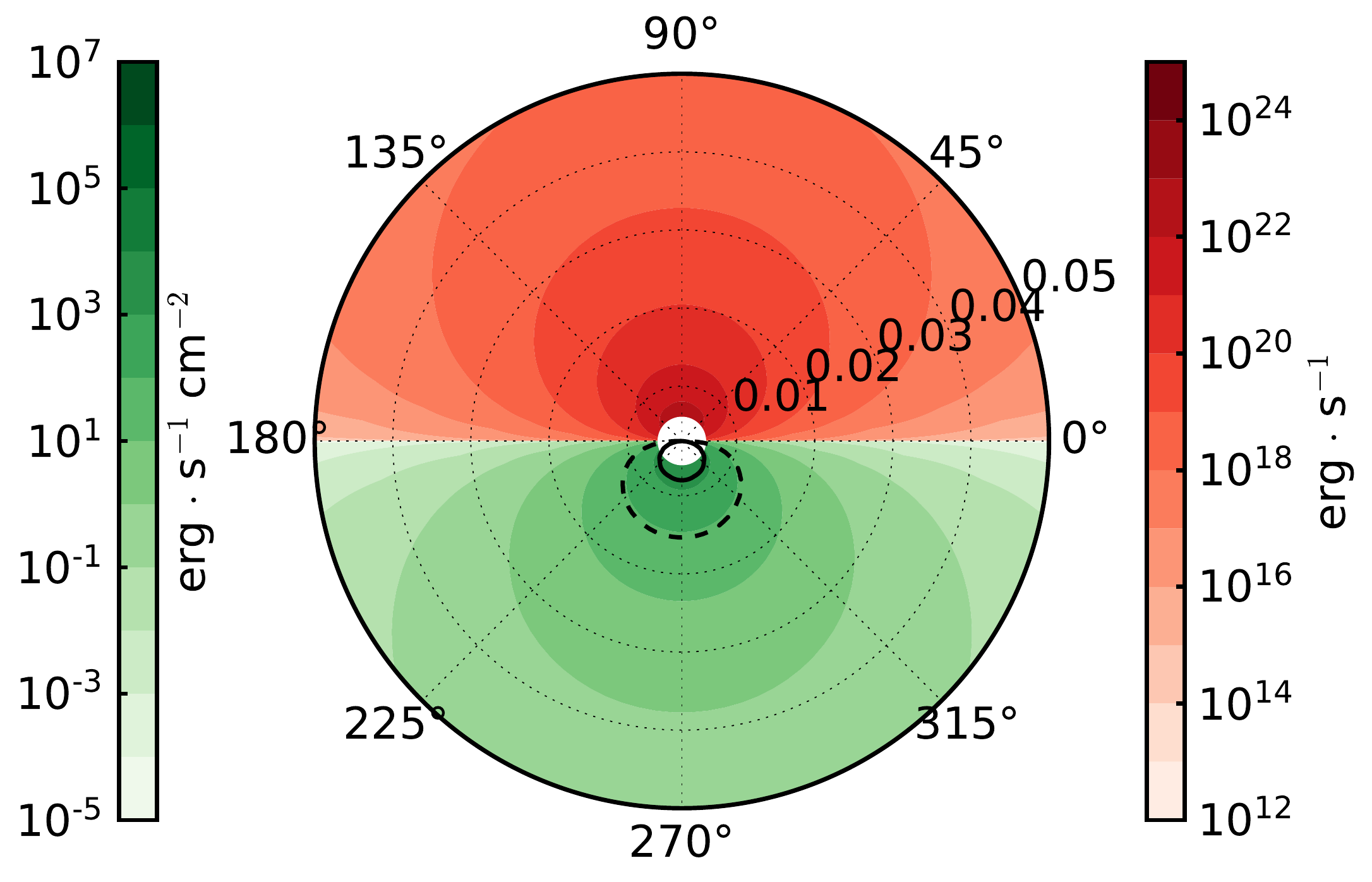}       
        \caption{Total energy release inside an Earth-radius and Earth-mass planet (upper half, in red) and surface heat flux (lower half, in green) as a function of distance to WX~UMa (concentric dotted rings) and orbital inclination. The solid and dashed black lines mark the surface heat fluxes of $2\times10^3$\,erg\,cm$^{-2}$\,s$^{-1}$, which corresponds to the surface heat flux on Io due to tidal heating, and 80\,erg\,cm$^{-2}$\,s$^{-1}$, which corresponds to the total heat flux of modern Earth. The white area in the center shows the unstable region inside the Roche limit of the star. Upper left: energy release and heat flux assuming an observed dipole magnetic field of 7.3\,kG calculated for an Earth-like conductivity profile. Lower left: same as upper left but for a lower magnetic field of 3\,kG. Right: same as the left panels, but for a molten mantle with a higher conductivity of $5\times10^{10}$\,CGS ($\approx 5.6$\,Sm\,m$^{-1}$). For comparison, modern terrestrial level of energy release due to radioactive decay is $\approx2.1\times10^{20}$~erg~s$^{-1}$. In the past (4.5~Gyr ago), energy release due to radioactive heat sources has been higher and equal to $\approx9.3\times10^{20}$~erg~s$^{-1}$ \citep{Schubert01}.}        
     \label{f_energy}
  \end{center}
\end{figure*}

Figure~\ref{f_energy} presents our main results. The four panels show energy release and surface heat flux in the planetary mantle depending on the orbital separation and inclination, and further considering two different stellar dipolar magnetic field strengths and two planetary conductivity profiles. For both magnetic fields, there is a region close to the star where the surface heat flux due to induction heating exceeds $2 \times 10^3$\,erg\,s$^{-1}$\,cm$^{-2}$ (2\,W\,m$^{-2}$) by up to two orders of magnitude. This value corresponds to Io's heat flux induced by tidal heating, which makes this Jovian satellite the most volcanically active body in the solar system. As expected, the energy release is lower for the lower stellar magnetic field strength and the parameter space within which the surface heat flux exceeds 2\,W\,m$^{-2}$ shrinks considerably, although it does not completely disappear. Even at larger orbital distances, induction heating is still more powerful than the modern Earth energy release due to radioactive decay.

\begin{figure*}
  \begin{center}
        \includegraphics[width=1.0\columnwidth]{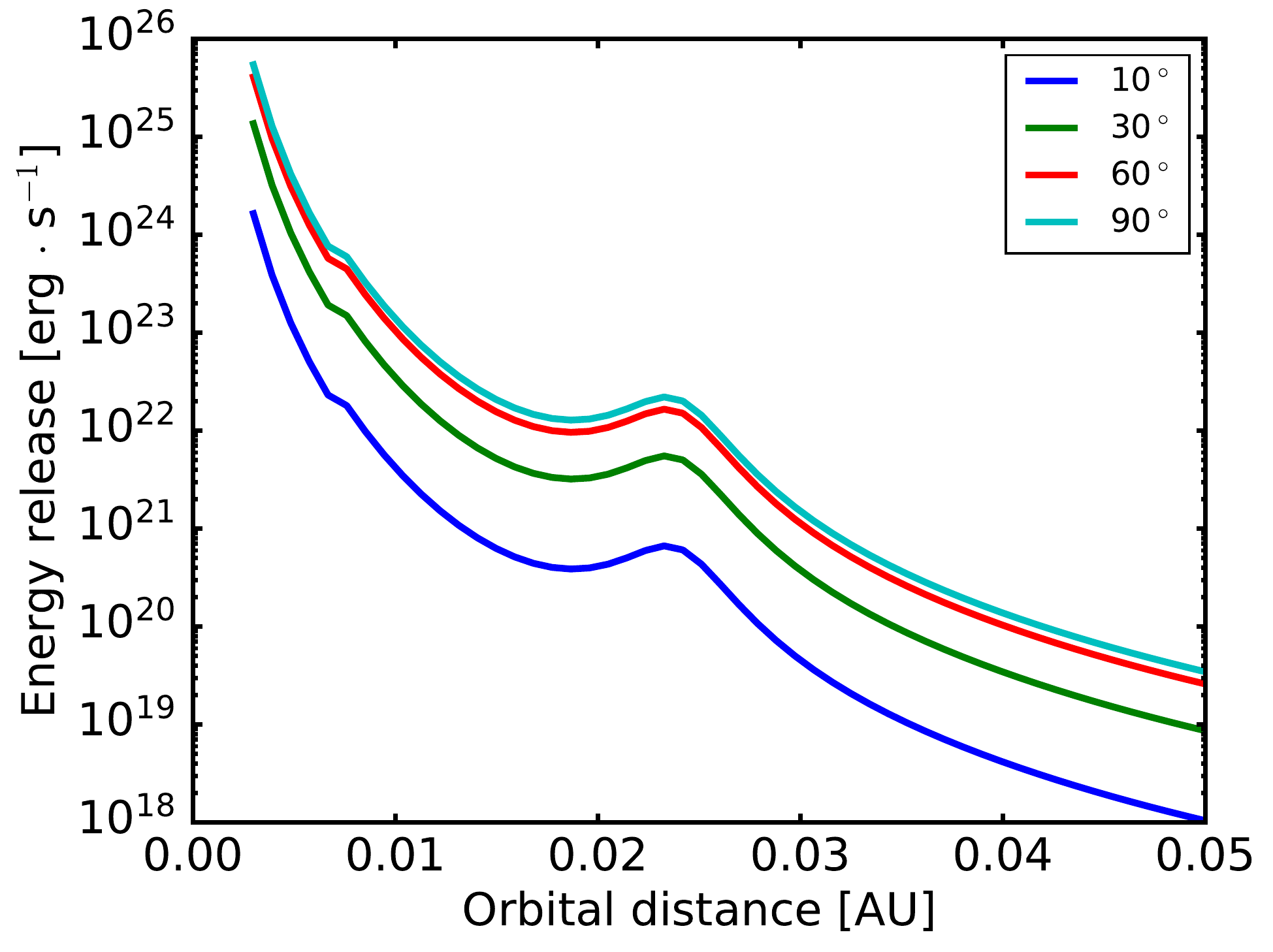}
        \includegraphics[width=1.0\columnwidth]{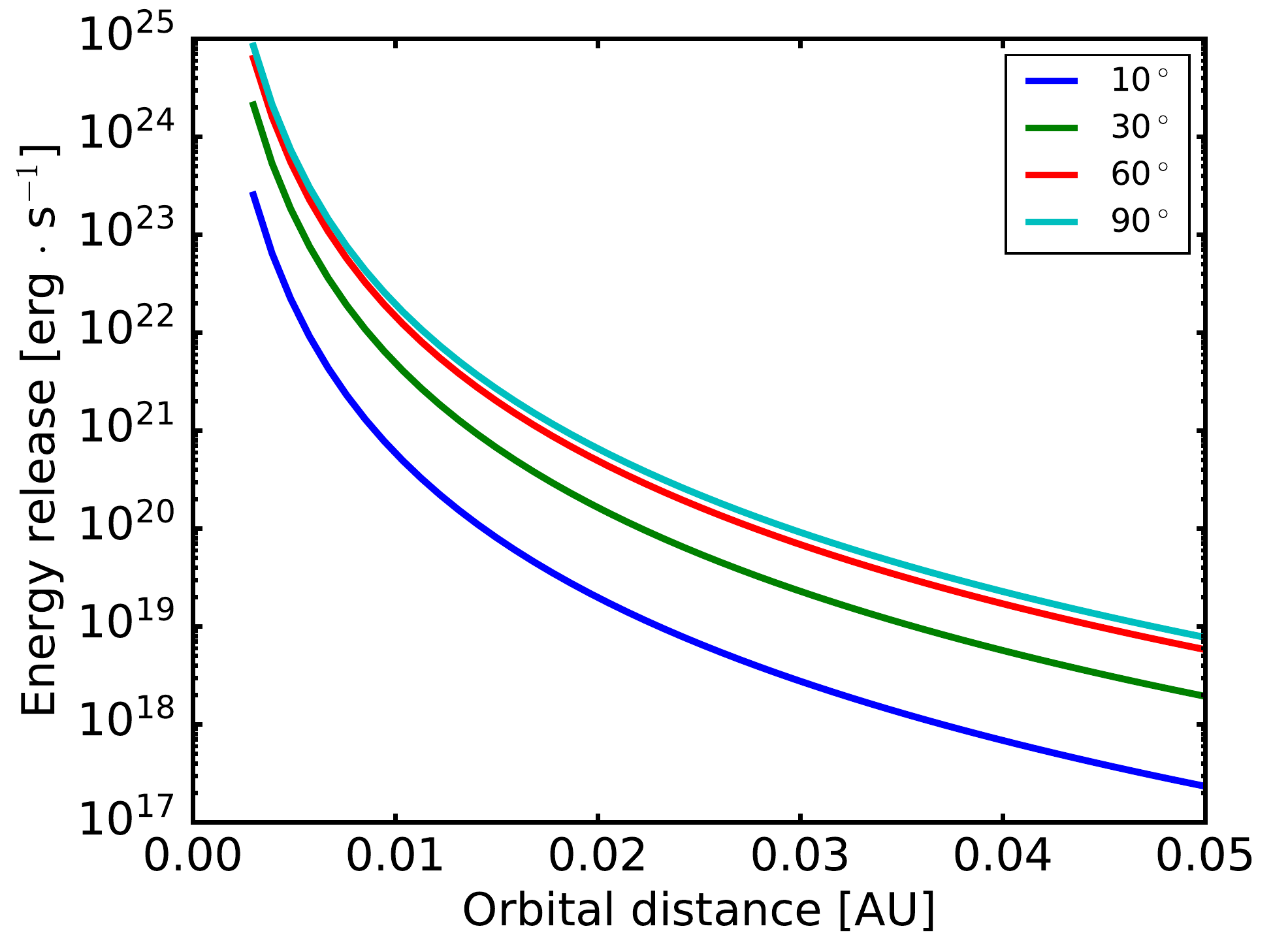}
        \caption{Total energy release as a function of orbital distance to WX~UMa for four different orbital inclinations and considering an Earth-like conductivity profile (left) or a molten mantle with a uniform conductivity (right).}        
     \label{f_orb}
  \end{center}
\end{figure*}

\begin{figure*}[t]
  \begin{center}
        \includegraphics[width=1.0\columnwidth]{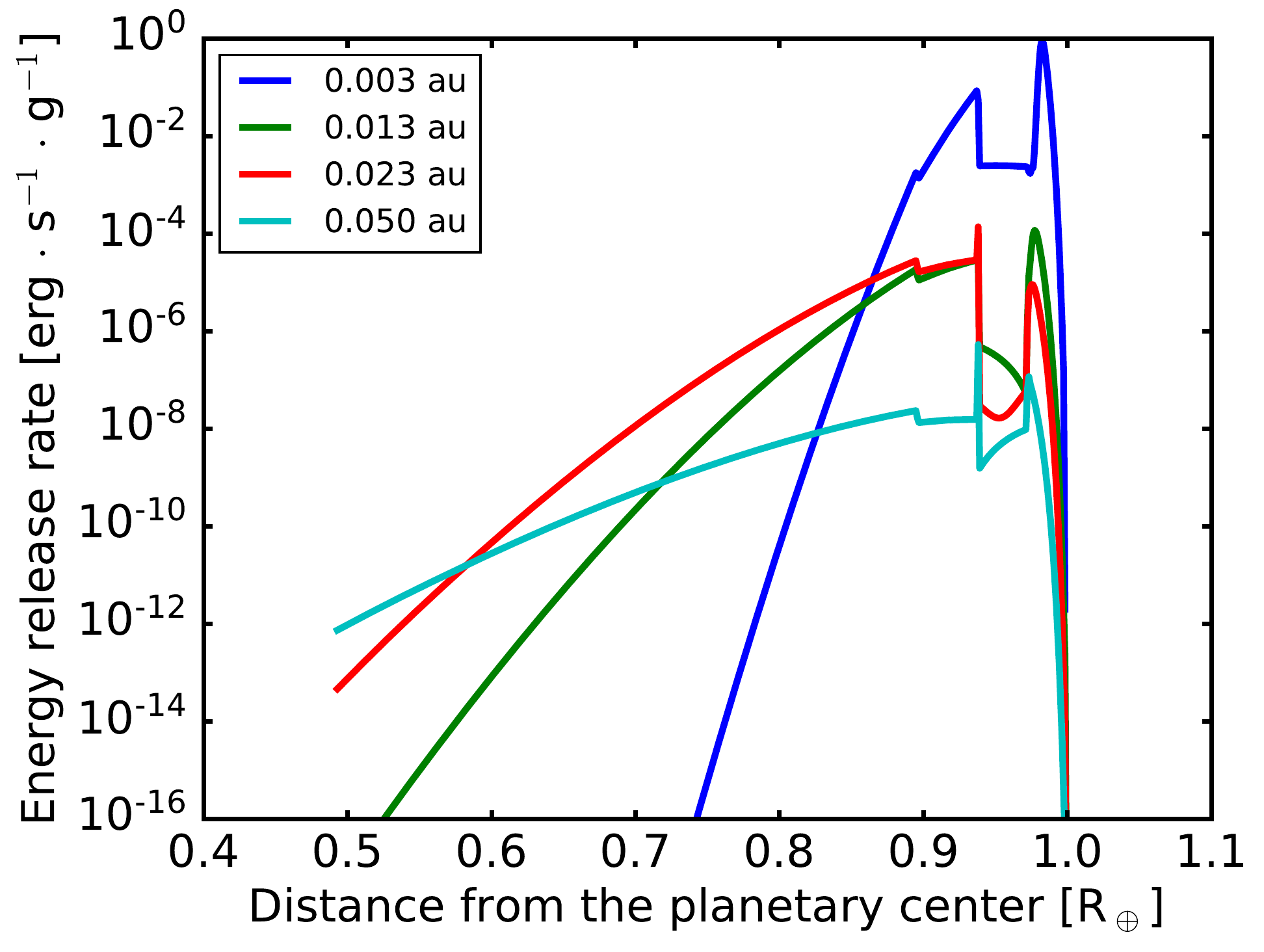}
        \includegraphics[width=1.0\columnwidth]{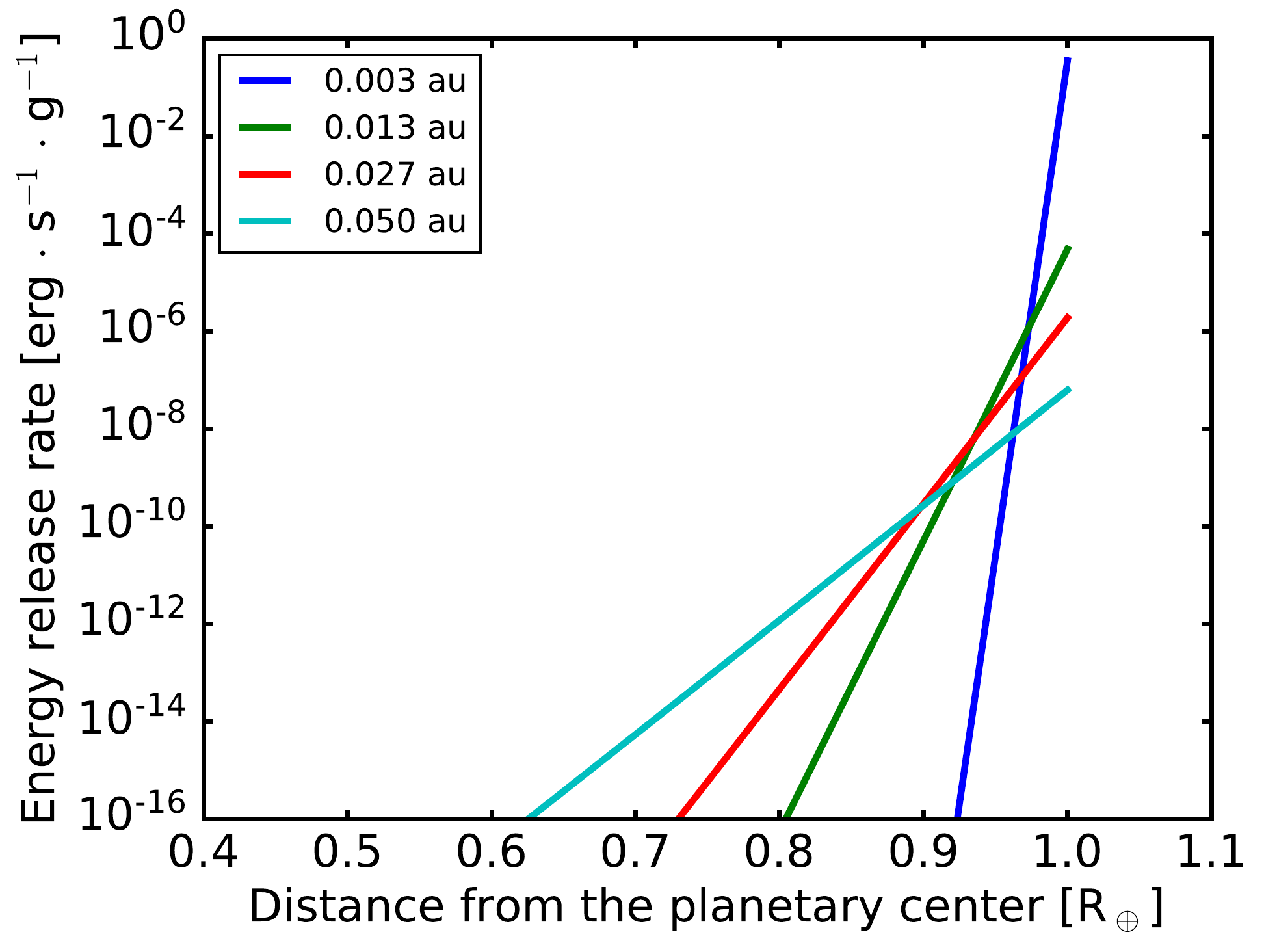}      
        \caption{Energy release inside the planet for the Earth-like (left) and uniform (right) conductivity profiles for different orbital distances. Both pictures show energy release for the orbital inclination of 90$^\circ$ and stellar magnetic dipole field of 7.3~kG. For the Earth-like profile, at 0.023~au a local maximum of energy release is located. One can see that in comparison to 0.013~au, energy can be released inside a bigger volume and the location of the maximum energy release is shifted deeper into the planetary mantle. This is due to more effective penetration of the magnetic field inside the planetary mantle for lower frequency. For a uniform conductivity profile, energy release is monotonically decreasing with increasing orbital distance. }        
     \label{f_mass}
  \end{center}
\end{figure*}

The high internal heat fluxes found for most of the parameter space considered here indicate that the planetary mantle would be fully molten within geologically short time scales. Jupiter's satellite Io possesses a molten mantle caused by tidal heating despite its comparable or even lower energy release rate. It seems therefore reasonable to expect the rocks in the interior of a planet around WX~UMa to be molten. 

It is possible that tidal heating preheats the mantle, so that induction heating has a somewhat smaller impact because molten rocks have higher conductivities than their solid counterparts. The right panels of Fig.~\ref{f_energy} illustrate the strength of induction heating for a planet with a solid surface and a molten mantle, again for for stellar magnetic field strengths of 7.3 and 3\,kG. Induction heating decreases with increasing conductivity because the magnetic field is screened more efficiently and does not penetrate so deep into the planetary mantle. However, it is still very substantial even for the molten mantle case. 

One can imagine that planets with orbital separations between 0.01 and 0.02\,AU have a mantle with some high melt fraction, however not completely molten, so that energy release due to induction heating would balance energy losses due to volcanic activity and radiation to space. Detailed interior modeling of these processes is beyond the scope of the present study.

The white area at the center of each panel in Fig.~\ref{f_energy} indicates orbits lying inside the Roche limit (tidal disruption radius) calculated as \citep{Agol11}
\begin{equation}
a_R = 0.0054 \left(\frac{\rho_{pl}}{\rho_\oplus}\right)^{-1/3} \left(\frac{M_{st}}{0.6 M_\odot }\right)^{1/3}\,,
\end{equation}
where $\rho_{pl}$ is the planetary mass density and $M_{st}$ is the mass of WX~UMa. Planets at orbits closer than the Roche limit are torn apart by tidal forces.

In the habitable zone, which for late M~dwarfs is close to 0.05\,AU, induction heating can still influence the interior evolution of a planet by increasing outgassing and volcanic activity; however, it is unable to produce a fully molten mantle for most mantle parameters (K17). The left panels of Fig.~\ref{f_orb} (showing the case of the Earth-like conductivity profile) show that the decrease of the energy release is non-monotonic. This is caused by an interplay between the penetration depth of the external magnetic field and the decrease of the magnetic field at the planet's orbit with an increasing distance to the star. While the magnetic field is monotonically decreasing with distance, the penetration depth increases with decreasing frequency of the magnetic field variation (the latter is assumed to be equal to the frequency of the orbital motion, which is lower further away from the star). Therefore, for some conductivity profiles, induction heating may be higher for slightly lower magnetic field, but for a larger penetration depth (lower frequency). Fig.~\ref{f_mass} shows the distribution of the heating rate inside the planet for both conductivity profiles. A high frequency of magnetic field variation increases the energy dissipation in a given volume, but it also leads to a very fast decline of the magnetic field inside the planetary mantle. On the other hand, a lower frequency corresponds to less dissipation per unit of volume, but it also allows the field to penetrate deeper inside the planet, which increases the volume where energy can be dissipated. For this reason, energy release at 0.023~au exceeds energy release at 0.013~au, despite slightly lower magnetic field at the latter orbital distance (see left panel of Fig.~\ref{f_mass}). For a fixed conductivity of $5 \times 10^{10}$\,CGS ($\approx 5.6$\,Sm\,m$^{-1}$), the increase of energy release at $\approx$0.025\,AU disappears (Figs.~\ref{f_orb} and \ref{f_mass}, right panels). If one would further increase the frequency, one would see a decline in energy release also for the uniform conductivity profile, but such high frequencies can not be reached from the planetary orbital motion.

\subsection{Orbital decay due to induction heating}

Energy which supplies the induction heating is derived from the orbital motion. In this subsection, we check if energy release due to induction heating is sufficient for a rapid orbital decay.

To do this, we follow the approach by \citet{Laine08} and compare the released energy to the total energy of the Keplerian orbit. At a given distance, this energy equals
\begin{equation}
E_i = - \frac{G M_{pl} M_{st}}{2 R_{i}},
\end{equation}
where $G$ is the gravitational constant, $M_{pl}$ is the planetary mass equal to one Earth mass, $M_{st}$ is the mass of WX~UMa, and $R_{i}$ is planetary orbital distance. To calculate the characteristic time it would take to evolve the orbit from one position to another, $\tau(R_{i})$, we calculate the difference of the potential energy between two orbital locations (at $R_{i}$ and $R_{i-1}$) and compare it to the energy release due to induction heating at this orbital distance which is shown in Fig.~\ref{f_orb}. The total time it takes for the orbit to decay from the starting point to the tidal disruption radius is then calculated as $\sum_i \tau(R_{i})$ for all $R_{i}$. We perform all calculations for a stellar dipole field of 7.3~kG and an orbital inclination of 90$^\circ$, i.e., for the maximum possible energy release.

\begin{figure*}
  \begin{center}
        \includegraphics[width=1.0\columnwidth]{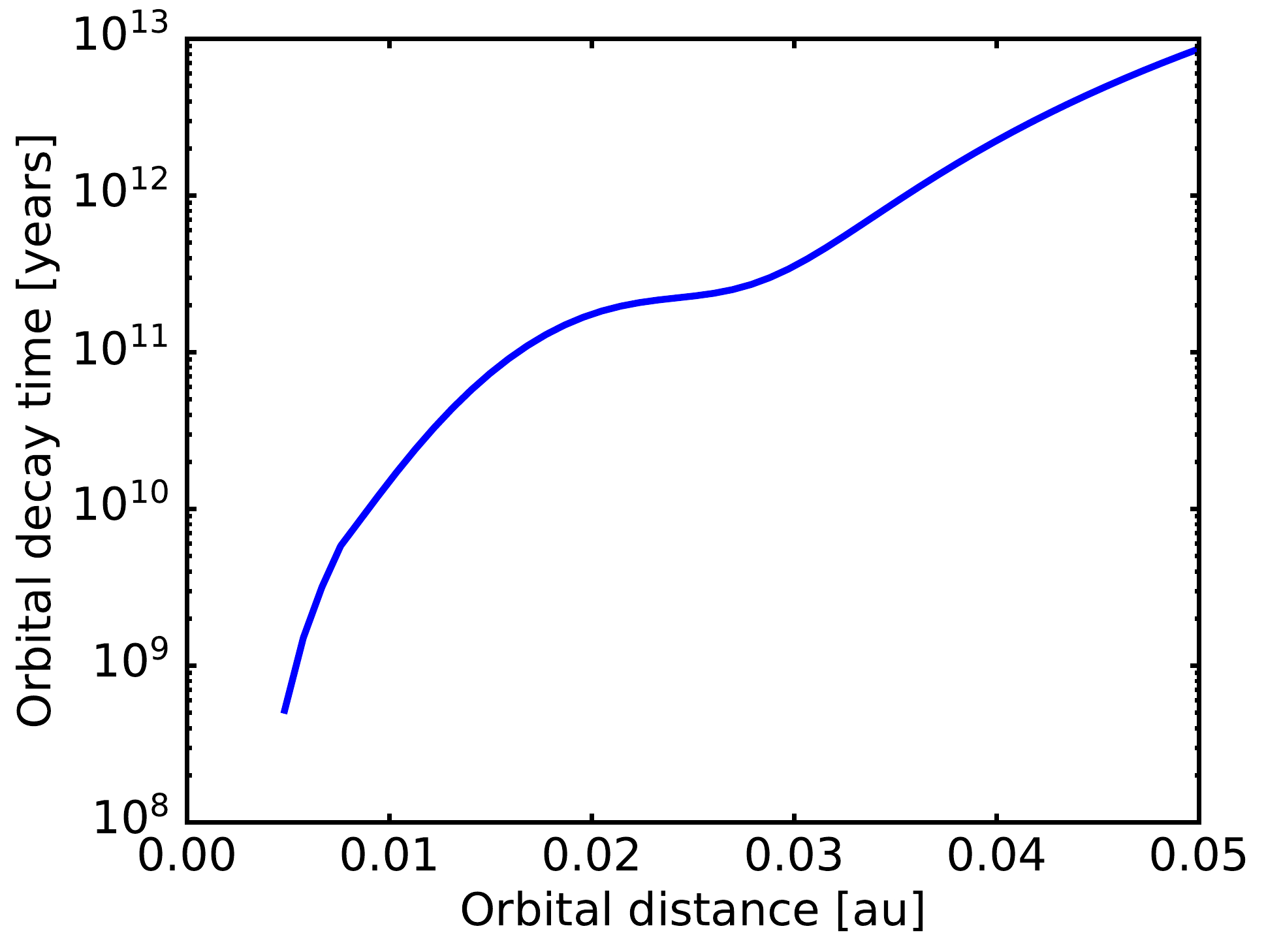}
        \includegraphics[width=1.0\columnwidth]{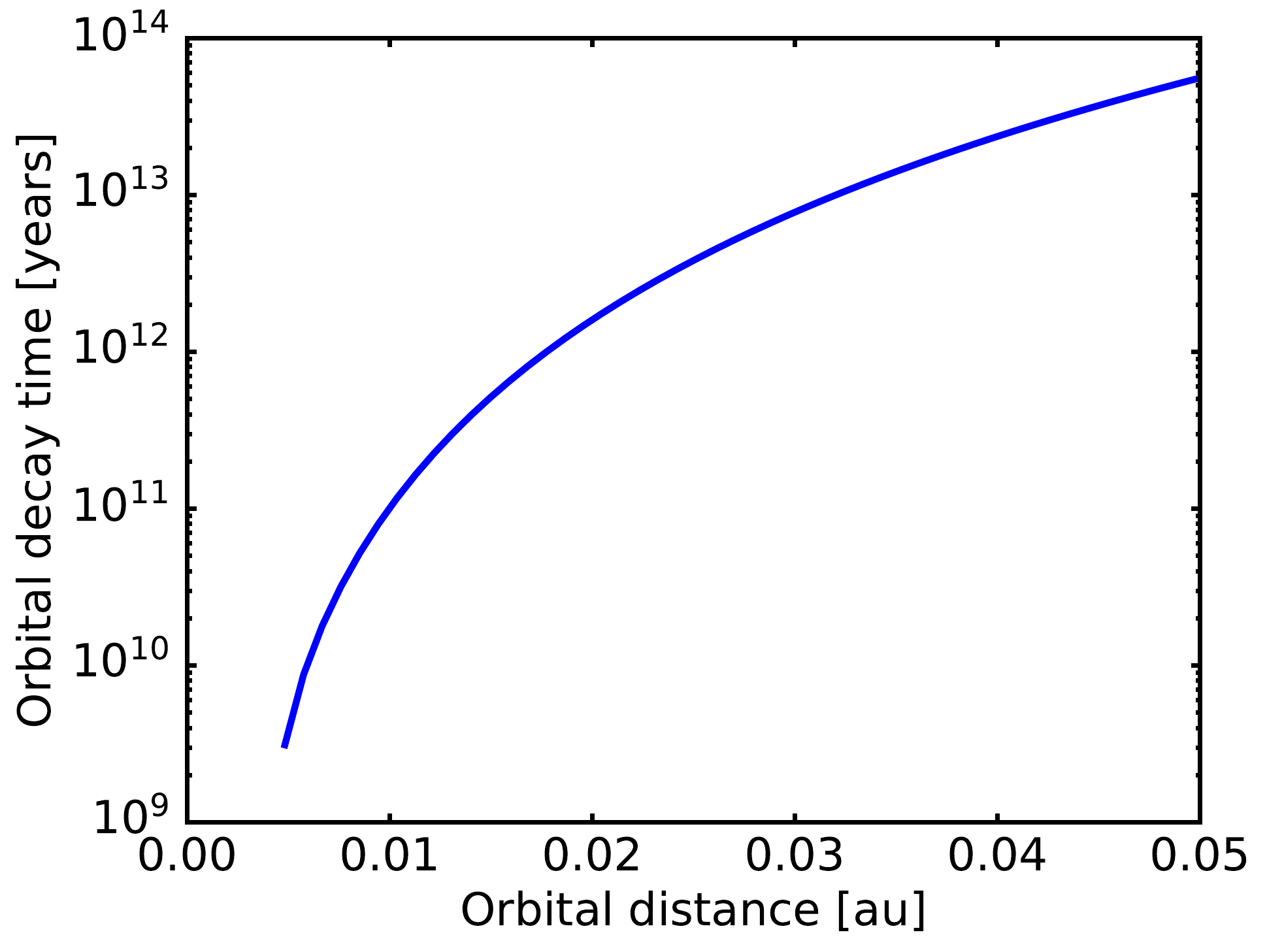}
        \caption{Characteristic time to degrade the orbit to the tidal disruption point for the Earth-like conductivity profile (left) and the case of constant conductivity (right) for a given initial orbital distance (x-axis).    }    
     \label{f_tau}
  \end{center}
\end{figure*}

The total time of orbital decay obviously depends on the initial orbital distance. Fig.~\ref{f_tau} illustrates how long it takes for the orbit to degrade due to energy dissipation caused by induction heating depending on the initial orbital distance. Since the planetary mantle will likely be molten within a geologically short time, the orbital evolution time for the molten mantle presents a more realistic estimate. As one can see, orbital decay due to induction heating is not very fast for most orbital distances. This conclusion is similar to the estimate by \citet{Laine08}, who has shown that energy release due to induction heating is not sufficient to stall the planetary migration of a hot Jupiter orbiting a T~Tauri star at 0.04~au. However, depending on the conditions in the system, a planet might still migrate due to other mechanisms, such as tidal or unipolar inductor effects (e.g., \citealp{Strugarek17}).

\section{Observability}
\label{sec_obs}

Induction heating can lead to high volcanic activity and, accompanied by atmospheric escape, form a plasma torus along the planetary orbit, similar to Io's plasma torus. Io's volcanoes mostly produce SO$_2$ molecules, which then dissociate to form S and O atoms. Volcanoes can also expel CO$_2$, however, volcanic activity depletes CO$_2$ in the mantle (as is the case for Io), until volcanoes cease to outgas it. Therefore, one can expect S and O atoms in the plasma tori of exoplanets strongly heated by induction, as well as C for the younger planets. 

\begin{figure}
  \begin{center}
        \includegraphics[width=1.0\columnwidth]{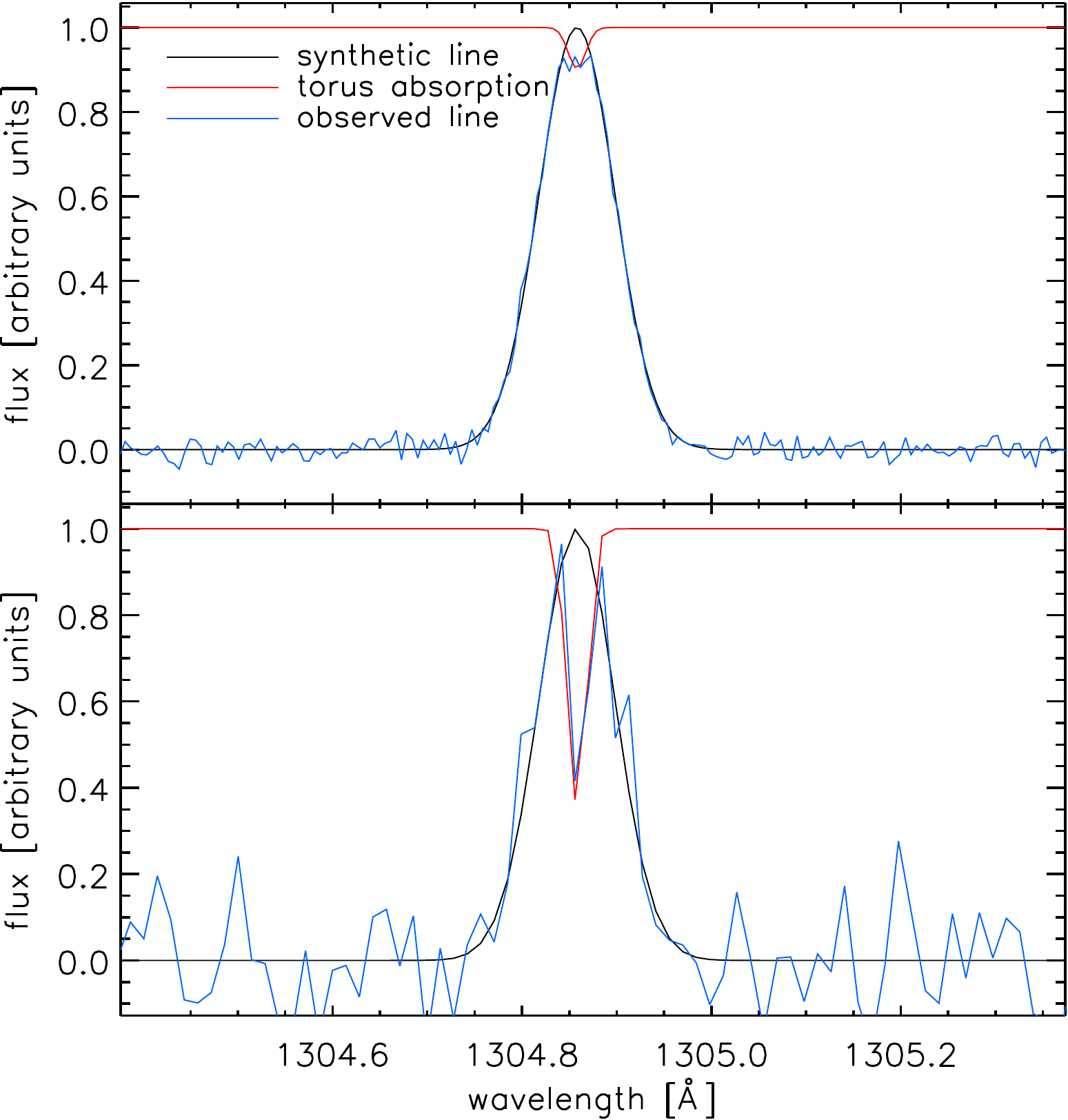}
        \caption{Synthetic Gaussian profile of the O{\sc i} line at 1304.858\,\AA\  (black) superposed to a synthetic Voigt absorption feature (red) computed assuming a column density of 10$^{12}$\,cm$^{-2}$ (top) and 10$^{13}$\,cm$^{-2}$ (bottom). The emission lines including the torus absorption (blue) were computed assuming a signal-to-noise ratio of 50 (top) and 10 (bottom). We employed the spectral scale and resolution of the E140H (top) and E140M (bottom) HST/STIS gratings. The broadening of the emission lines is dominated by the intrinsic stellar line broadening of 0.08\,\AA\  (full-width-half-maximum), derived from the HST/STIS spectrum of AD\,Leo, while for the absorption feature we employed a broadening $b$-parameter of 3\,km\,s$^{-1}$.}
     \label{fig:observability}
  \end{center}
\end{figure}

A dense-enough plasma torus can absorb the stellar light at the position of strong resonance lines of abundant elements, as in the case of the WASP-12 system \citep{Haswell12,Fossati13,Kislyakova16}. To estimate if the presence of such torus could make the effects of combined effect of induction, tidal and radioactive heating indirectly detectable, we consider the detectability of absorption signatures by O{\sc i} superposed to the stellar far-ultraviolet (FUV) emission triplet at $\lambda\approx$\,1304\,\AA. Oxygen is one of the most abundant elements escaping the planetary surface; it is hardly ionized by the stellar radiation \citep[e.g.,][]{Mura11}, and the FUV O{\sc i} triplet is prominent in the FUV spectrum of cool stars. We produce synthetic O emission lines taking the HST/STIS E140M observations of the active M~dwarf AD~Leo as reference for the line strength and width. We then place absorption features on top of the emission lines assuming a uniformly dense torus covering the whole stellar disk, the best case scenario of the absence of interstellar medium absorption (this would however affect mostly the 1302\,\AA\ feature, which is the only resonance line in the triplet), and the ideal case of no velocity displacement between the emission and absorption features. We find that the torus should have an O column density of 1--3$\times$10$^{12}$\,cm$^{-2}$ to be detectable on a spectrum with a signal-to-noise ratio of 50 obtained with an instrument grating similar to HST/STIS E140H (spectral resolution of 114\,000; Fig.~\ref{fig:observability}, top). Such a high data quality would be reachable for most nearby M-dwarfs with instruments such as LUMOS/POLLUX on board LUVOIR \citep{France17}. For a more moderate data quality, such as a spectral resolution of about 50\,000 (e.g., HST/STIS E140M grating) and a signal-to-noise ratio (S/N) of 10, which is currently reachable for nearby M-dwarfs with HST, the torus would need to have a column density larger than 10$^{13}$\,cm$^{-2}$ (Fig.~\ref{fig:observability}, bottom).

To see whether such a situation could be possible, we employed Eq.~7 of \citet{Haswell12}, which gives
\begin{equation}
N(O) \leq \frac{\dot{m}}{2 \pi v R_{\star} \mu}\,,
\label{eq:NO}
\end{equation}
where $N(O)$ is the O{\sc i} column density, $\dot{m}$ is the mass-loss rate of O atoms, $v$ is the velocity at which the particles in the torus leave the system, $R_{\star}$ is the stellar radius, and $\mu$ is the mean molecular weight of the gas in the torus. We set $N(O)$ equal to 10$^{12}$\,cm$^{-2}$ and assume that the torus is composed mostly by O{\sc i}, thus setting the value of $\mu$. It is not possible to estimate the O mass-loss rate from our computations, therefore we assumed it to be equal to that of Io \citep[2$\times$10$^{28}$\,g\,s$^{-1}$][]{Thomas04}. This leaves $v$ as the only fully unconstrained parameter in Eq.~\ref{eq:NO}. Therefore, we estimated the maximum velocity that is required to maintain a torus column density high enough to be detectable, obtaining a value of the order of 10\,km\,s$^{-1}$, which seems to be plausible considering that O atoms are only weakly affected by radiation pressure \citep[e.g.,][]{Mura11} and, being neutral, are not efficiently picked up by the stellar wind. The maximum value of $v$ decreases then to about 1\,km\,s$^{-1}$ if one considers that, because of the different surface areas, a 1\,R$_{\oplus}$ planet would have a mass-loss rate that is about 10 times larger than Io's. These order of magnitude estimates of the possible detectability of a torus produced by O{\sc i} escape from a volcanically active planet are independent on the source of the mechanism heating the planet's interior.

 We further estimated the number of M dwarfs for which it would be possible to look for the torus absorption signature with HST. We considered as reference the O{\sc i} triplet observed for AD\,Leo and made use of the STIS exposure time calculator to estimate the stellar flux required to reach a S/N of 10. For this exercise, we considered the STIS G140M grating with the 1321 setting. We concluded that, within an exposure time corresponding to three HST orbits (about 9000 seconds), it would be possible to detect the torus absorption for AD\,Leo-like M-dwarfs lying as far as about 46\,pc. Extending the exposure time to ten HST orbits (about 30000 seconds), the radius within which it would be possible to detect the torus absorption extends to about 65\,pc. Detecting the torus absorption for stars at larger distances would become instrumentally challenging, even by significantly increasing the exposure time. This is because of the rather high background level of the STIS far-UV detector. Deeper observations would therefore require the use of an instrument with a significantly lower background, such as COS or those designed for LUMOS/POLLUX on board LUVOIR.

All strongly magnetic M-dwarfs for which the magnetic field has been measured and mapped (roughly 20--30) lie within this distance \citep[e.g.,][]{Morin10,Shulyak17}. Considering the large number of M-dwarfs present in the solar neighborhood and that the vast majority of the fully convective M-dwarfs are expected to be strongly magnetic \citep{Yadav15}, our results indicate that there should be hundreds of late-type M-dwarfs for which the detection of the torus absorption would be, in principle, possible. However, the presence of a detectable absorption would first require that a star hosts a close-in rocky planet and that the orbital geometry is such that the torus lies along our line of sight. In addition, the detection of a torus is made challenging, particularly with HST, by the presence of ISM absorption, time variability in the stellar emission features, and velocity shifts between the stellar emission and torus absorption features. Future facilities should be able to reach the required high-quality observations, particularly in the far-UV (e.g., LUVOIR), and also allow the detection of planets orbiting fully convective M-dwarfs, as well as measure and characterise their magnetic fields \citep[e.g., with the SPIRou spectropolarimeter;][]{spirou}.

\section{Conclusions}
\label{sec_consclusions}

We apply the model developed by K17 to an Earth-like planet orbiting WX~UMa. This star has one of the strongest magnetic fields observed on a low-mass main-sequence stars, which is dominated by the dipole component. Since the dipole and rotational axes of this star are co-aligned, the planet experiences a constantly varying magnetic field only if its orbit is inclined. For very close orbital distances, the unipolar inductor model (a Jupiter-Io-like interaction, where a magnetic flux tube is formed connecting the two bodies; see, e.g., \citealp{Goldreich69,Laine12,Buzasi13}) may be applicable together with the induction model, but considering it is beyond the scope of the present study.

We show that for some inclinations and close orbital distances, energy release due to induction heating is so high that it exceeds the surface heat flux of Io, the most volcanically active body in the solar system. From the observations and interior models, we know that energy releases of such magnitude lead to the formation of a magma ocean beneath the solid surface. Induction heating is strong also for planets with a molten mantle. Therefore, it is likely that the planets on inclined orbits around WX~UMa-like stars, or even M~dwarfs with weaker magnetic fields, may experience extreme volcanism and the possible formation of a plasma torus along their orbits. For young planets, the outgassed material may contain a large fraction of CO$_2$, while later, when the mantle is already somewhat depleted of volatiles, an Io-like composition of the torus is possible. This torus could be observed in O{\sc i} lines. 
We conclude that induction heating can be a very powerful energy source for rocky planets orbiting strongly magnetized M~dwarfs and should be taken into account among other heating sources when addressing the interior evolution of such planets.

\acknowledgements{We acknowledge the support by the Austria Science Fund (FWF) NFN project S116-N16 and the subprojects S11607-N16 and S11604-N16. TL acknowledges funding via the Austrian Space Application Programme (ASAP) of the Austrian Research Promotion Agency (FFG) within ASAP11. The authors thank D.~Shulyak and S.~Boro~Saikia for the discussion about the newest measurements of the stellar magnetic fields and R.~Yadav for the discussion on the dynamo modeling of fully convective stars.}


\end{document}